\documentclass[12pt,preprint]{aastex}  
\shortauthors{Fassnacht \& Taylor}
\shorttitle{CSOs as Flux Density Calibrators}

\newcommand{\rahr}{$^{\rm h}$}
\newcommand{\ramn}{$^{\rm m}$}

\begin{document}

\slugcomment{Accepted to the Astronomical Journal}
\title{~~\\~~\\ {Compact Symmetric Objects as \\ Radio Flux Density Calibrators}}

\author{
        C. D. Fassnacht\altaffilmark{1}
        and G. B. Taylor
}
\affil
{
        National Radio Astronomy Observatory
}
\affil
{
        P. O. Box O,
        Socorro, NM 87801
}
\email {
        cdf@stsci.edu,
        gtaylor@nrao.edu
}
\altaffiltext{1}
{
        Current Address:  
        Space Telescope Science Institute
        3700 San Martin Drive
        Baltimore, MD 21218
}

\begin{abstract}

We present results from the first intensive monitoring campaign of a
sample of Compact Symmetric Objects (CSOs).  We observed seven CSOs at
8.5~GHz over a period of eight months, with an average spacing between
observations of 2.7 days.  Our results show that, as predicted, the
flux densities of the CSOs are extremely stable; the mean RMS
variability of the sample was 0.7\% in flux density.  The low
variability of the CSOs makes them excellent flux density calibrators
at this frequency.  We recommend that at least four CSOs be included
in any VLA monitoring campaign which requires precise epoch-to-epoch
calibration, such as those to measure gravitational lens time delays.
The CSO data enable the correction of small systematic errors in
the primary flux calibration.

\end{abstract}


\section{Introduction}

For calibration of interferometric observations, it is generally
desirable to employ unresolved calibrator sources.  These sources have
the advantage of appearing the same to all antenna pairs
(visibilities), regardless of baseline length or orientation.
Ideally, one would also prefer the flux density of the calibrators to
be constant in time in order to facilitate removal of systematic
atmospheric or instrumental variations.  Calibrators that are
unpolarized have added benefits in solving for the instrumental
polarization terms.

Unfortunately (from the perspective of calibration) most compact (size
$<$ 0\farcs1) extragalactic radio sources are highly variable.  These
sources, which have flat radio spectra, have flux densities that vary
by $\sim$10\% on timescales of weeks to months and can easily vary by
$\sim$100\% on timescales of months to years \citep{aller90}.
\citet{qui92} found low-amplitude (1--2\%) intra-day variability in
all flat-spectrum sources, and larger variations (up to 20\%) in
$\sim$25\% of these sources.  VLBI imaging of a large flux-limited
sample \citep{tay1996} reveals that 95\% of compact, flat-spectrum
sources have asymmetric, core-jet morphologies.  The core is generally
the dominant component and has a typical size of less than 0.1 mas
(under a parsec for typical redshifts).  The variability can thus be
explained by the short light-crossing time of the emitting volume,
and/or refractive interstellar scintillation in the Galactic ISM.  The
jet components exhibit superluminal motions and are probably moving at
relativistic speeds \citep{coh94}.  Flares in the jet components have
also been observed with similar amplitudes and timescales as core
flares \citep{zhou2000}.

There is, however, a small fraction \citep[5\%;][]{csosamp} of compact
sources whose pc-scale structure is dominated by steep-spectrum
extended emission on both sides of the core. These sources are known
as Compact Symmetric Objects (CSOs).  Recent measurements of kinematic
ages of $\sim$1000 years \citep{ows98,owspol98,tay2000} support the
theory that CSOs are small by virtue of their youth and not because
they are frustrated by a dense environment.  This was the favored
interpretation by \citet{phi80,phi82} who first drew attention to a
group of compact double sources with steep or GHz-peaked spectra and
slow motions compared to the majority of core-jet sources.  At about
the same time, \citet{rud82} reported that GHz-peaked spectrum (GPS)
sources generally have low variability and low polarization.  Although
there is some overlap between sources classified as CSOs and those
classified as GPS sources, we emphasize that there is no clear
correspondence between the two types of sources.  In fact, a recent
survey of 47 GPS sources found that only 3 could be clearly classified as
CSOs \citep{snel00}.  The term GPS is purely a spectral classification
while CSO is a physical one, requiring considerably greater
observational effort (usually VLBI observations at multiple
frequencies) in order to identify the location of the core component
and extended jet and/or lobe emission on both sides of the core.
Objects with vastly different powers, orientations, and evolutionary
stages can be classified as GPS sources, while the CSOs are a much
more homogeneous group.  The lack of correspondence between GPS and CSO
sources is especially important to keep in mind if one is choosing a
sample of CSOs as flux density calibrators.

The CSOs have qualities which should make them excellent flux
calibrator sources.  One is the relative unimportance of the
presumably variable core component.  For 6 bright CSOs,
\citet*{pinpoint} found the core to comprise from $<$0.4\% to 6\% of
the integrated flux density at 15 GHz.  Another quality is that the
jets and lobes of CSOs are thought to be relatively free of Doppler
boosting effects \citep{wil94}.  Thus, flux variations due to flares
in the jet components should not be magnified.  The physical
properties of CSOs have been summarized by \citet{rea96}, who pointed
out that the few CSOs known at that time exhibited both weak radio
variability ($<$ 10\%) and low polarization ($<$ 0.5\%).  In addition,
\citet{debruyn} reported that the CSO OQ\,208 \citep[classification
based on the VLBI images by][]{sta97} varied by less than 2\%~yr$^{-1}$
over a period of 10 years at 5 GHz.  \citet{csosamp} presented 8.4 GHz
VLBI observations showing that the polarization was less than 1\% in
21 CSOs and CSO candidates.  Yet another fortunate property of CSOs
for calibration purposes is that they generally exhibit no emission on
scales greater than 1 kpc.  An exception to this rule is the CSO
0108+388 which has some faint extended emission \citep{bau90}.
Another exceptional source is the CSO J1148+5924 (NGC~3894) which also
has faint kpc-scale emission and has been seen to be slowly variable
on timescales of years \citep{tay98}.  J1148+5924 is relatively nearby
at a redshift of 0.01085 \citep{jvg89}, and has a luminosity more than
two orders of magnitude less than all other members of the CSO class.

In \S2 we present observations of seven CSOs used as calibrators
in an intensive VLA monitoring campaign.  The calibration 
procedure is described in \S3.  In \S4 and \S5 we consider
primary and secondary systematic errors which allows us to provide, in \S6,
a prescription for obtaining high-precision monitoring.
Lastly, in \S7 we consider how variability studies can be used
to constrain physical models for CSOs.


\section{Observations}

The sample observed for this paper consisted of seven CSOs chosen from
the samples of \citet{csosamp} and \citet{tay1996}. These particular
sources were observed as part of a gravitational lens monitoring
campaign and were chosen because they were bright and were located in
the RA range defined by the lens systems.  The sources and their
coordinates are listed in Table~\ref{tab_obs}.  The observations were
made with the VLA\footnote{The National Radio Astronomy Observatory is
operated by Associated Universities, Inc., under cooperative agreement
with the National Science Foundation.} at 8.5~GHz during its A, BnA,
and B configurations, yielding typical angular resolutions of 0\farcs2
in the A configuration to 0\farcs7 in the B configuration.  The
monitoring program consisted of 88 observations of the sample between
1999 June 21 and 2000 February 14, giving an average spacing of 2.7~d
between epochs.  The typical observing block was one or two hours in
length, and the typical integration time on each of the CSOs was 45 or
60~sec.  For each epoch, some fraction of the CSO sample was observed
at a large hour angle, since the CSOs are widely spaced on the sky and
the observing blocks were short.  In fact, it was not always possible
to observe the entire sample of CSOs in an observing block.  However,
the incomplete sampling should not significantly affect the
conclusions of this paper.  All of the CSOs were observed at least 68
times, giving quite adequate determinations of their light curves.

\begin{deluxetable}{cccc}
\tablenum{1}
\tablewidth{0pt}
\scriptsize
\tablecaption{CSO Sample\label{tab_obs}}
\tablehead{
\colhead{Name (B1950)}
 & \colhead{Name (J2000)}
 & \colhead{RA(2000)}
 & \colhead{Dec(2000)}
}
\startdata
1031+567 & J1035+5628 & 10\rahr35\ramn07\fs040 & +56\degr28\arcmin46\farcs79 \\
1146+596 & J1148+5924 & 11\rahr48\ramn50\fs358 & +59\degr24\arcmin56\farcs38 \\
1242+410 & J1244+4048 & 12\rahr44\ramn49\fs196 & +40\degr48\arcmin06\farcs22 \\
1358+624 & J1400+6210 & 14\rahr00\ramn28\fs653 & +62\degr10\arcmin38\farcs53 \\
1543+480 & J1545+4751 & 15\rahr45\ramn08\fs530 & +47\degr51\arcmin54\farcs67 \\
1826+796 & J1823+7938 & 18\rahr23\ramn14\fs109 & +79\degr38\arcmin49\farcs00 \\
1946+708 & J1945+7055 & 19\rahr45\ramn53\fs520 & +70\degr55\arcmin48\farcs72 \\
\enddata
\end{deluxetable}

In addition to the CSOs and the lens systems, two phase calibrators
were observed as part of the monitoring program
(Table~\ref{tab_pcal}).  The typical integration times on the phase
calibrators were between 2 and 3~min.  These two sources will also be
discussed in this paper because they provide a useful contrast to the
CSO sample.  The lenses will be discussed in a series of future
papers.  All of the sources discussed in this paper can be found in
the list of VLA calibrators maintained by
NRAO\footnote{\url{http://info.aoc.nrao.edu/$\sim$gtaylor/calib.html}}.

\begin{deluxetable}{cccc}
\tablenum{2}
\tablewidth{0pt}
\scriptsize
\tablecaption{Phase Calibrators (Core-Jet Sources)\label{tab_pcal}}
\tablehead{
\colhead{Name (B1950)}
 & \colhead{Name (J2000)}
 & \colhead{RA(2000)}
 & \colhead{Dec(2000)}
}
\startdata
1547+507 & J1549+5038 & 15\rahr49\ramn17\fs469 & +50\degr38\arcmin05\farcs79 \\
1642+690 & J1642+6856 & 16\rahr42\ramn07\fs849 & +68\degr56\arcmin39\farcs76 \\
\enddata
\end{deluxetable}


\section{Calibration and Data Reduction \label{sec_calib}}

The data were calibrated in the AIPS software package developed by
NRAO.  The standard flux density calibrator 3C\,286 was not observed
for all epochs, so the overall flux density calibration was instead
tied to the source 3C\,343 (1634+628).  This source has a steep
two-point radio spectral index between 1.4 and 8.5~GHz ($\alpha \sim
-1.0; S_\nu \propto \nu^\alpha$) and its flux density has been shown
to be stable in past monitoring campaigns \citep{paper1}.  Although
the emission from 3C\,343 is dominated by the compact central
component, there is some low surface-brightness extended emission from
the source.  Because of the extended emission, 3C\,343 is not an ideal
flux calibrator.  First, the total flux density measured for the
source will change with the VLA configuration, since the more compact
configurations are more sensitive to the low surface-brightness
emission.  To correct for this effect, we scaled the 3C\,343 flux
densities by factors of 0.997 and 0.983 for the BnA and B
configurations, respectively.  Secondly, the changing $(u,v)$ coverage
for each monitoring epoch may cause small variations in the measured
flux density since different $(u,v)$ coverages are sensitive to
different source structures.  However, these small variations, as well
as those caused by the changing observing conditions and any intrinsic
variability in 3C\,343, can easily be corrected for, as shown in
\S\ref{sec_prim}.

Because the CSOs in our sample are compact and bright ($S_{8.5}>200$
mJy; Table~\ref{tab_rms}), they can be used to determine antenna gain
solutions without the need for an external calibrator.  For two
of the sources, J1035+5628 and J1400+6210, the emission from the source
is slightly resolved.  Thus, for these two CSOs, the gain solutions
were calculated only for baseline lengths that were less than
400~k$\lambda$.  For all of the other CSOs, all baselines were used.
For each CSO, two iterations of the AIPS task CALIB were run.  In the
first, only the gain phases were solved for, with a solution interval
of 10~sec.  The phase solutions were applied to the data and then the
second iteration of CALIB was performed.  In this iteration, both
amplitude and phase solutions were obtained, with a solution interval
of 30~sec.  The phase calibrators were processed in the same manner.

After the data were calibrated, we measured the CSO flux densities
using two separate methods.  The first method was to use the GETJY
task within AIPS, which computes the flux density based on the gain
solutions calculated in CALIB.  The second method was to
export the files from AIPS and to fit models to the $(u,v)$ data using
the {\tt difmap} package \citep{difmap}.  The {\tt difmap} processing
is more flexible in dealing with source structure than the GETJY task.
However, since the emission from CSOs is dominated by an unresolved or
barely resolved component, the two methods should not greatly differ
in the results that they produce.  This is, in fact what we see.  We
compared the GETJY and {\tt difmap} flux densities for all the CSOs
and found that the mean difference in flux density is
0.07\%$\pm$0.18\%.  The results presented in this paper are for the
{\tt difmap} flux densities.  However, we have also analyzed the data
obtained from the GETJY step and find no significant difference in
the results.  


\section{Correction for Primary Systematic Errors \label{sec_prim}}

\begin{figure}
\plotone{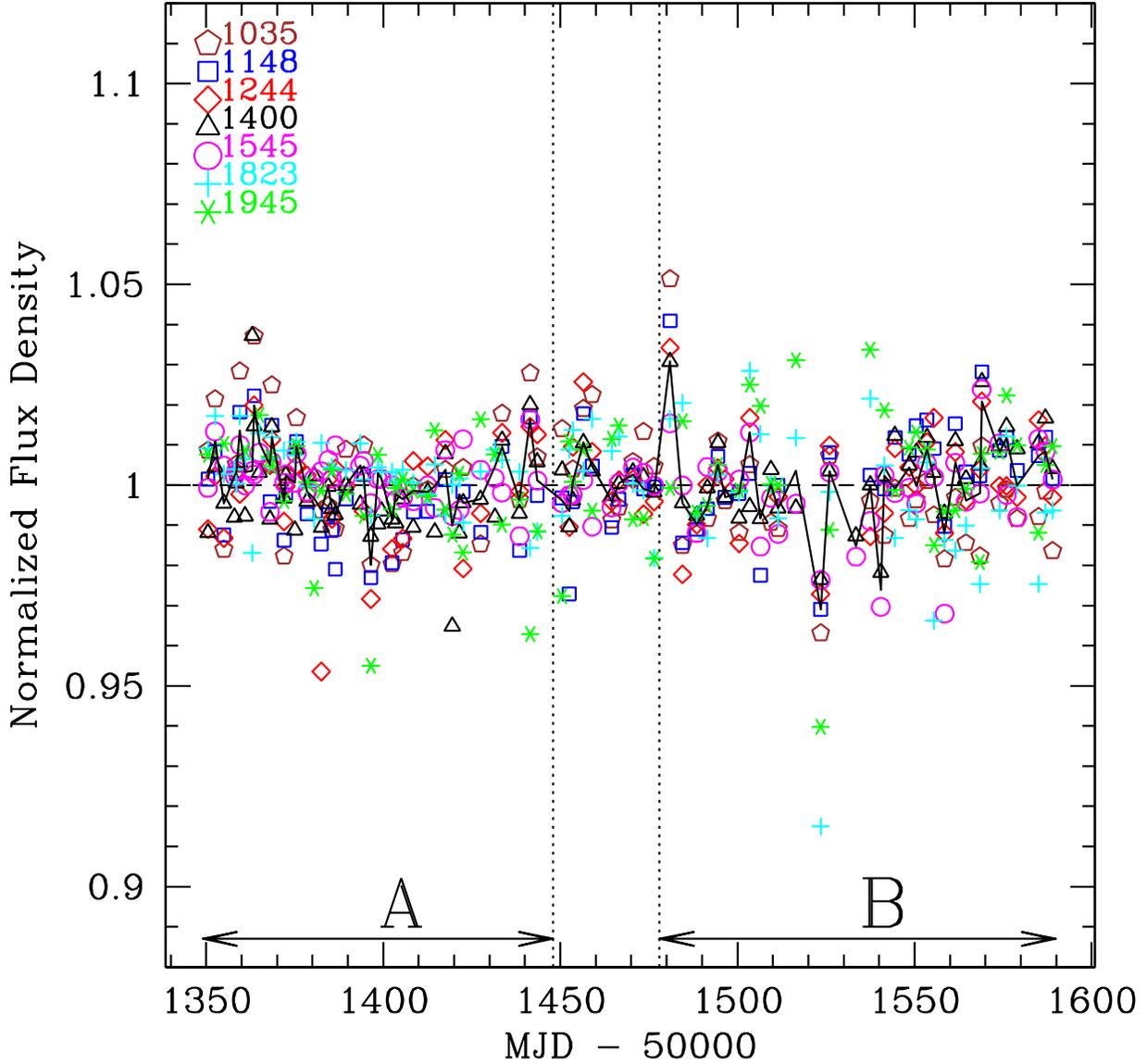}
\caption{Normalized light curves of the seven CSO sources.  The
abscissa is time, with units of MJD - 50000.  The ordinate is
normalized flux density, so the plotted points represent fractional
deviations from the mean.  The heavy black line indicates the median
normalized flux density, $M^{(1)}$, for each day of observation.  The
vertical dotted lines delineate the VLA configuration.  The A and B
configurations are labeled.  The unlabeled configuration is the BnA
configuration.  For clarity, the error bars on the points are not
shown.  The formal errors, estimated from the RMS noise in the
residual maps, are approximately the size of the points in the plot.
\label{fig_pltcal1}}
\end{figure}

To compare the variability of the CSOs in our sample, we normalized
the light curves of each CSO by the mean flux density, $<S_i>$,
recorded for that CSO over the course of the monitoring program.  The
values of $N_i$, the number of observations obtained on source $i$,
and $<S_i>$ are recorded in Table~\ref{tab_rms}.  We then defined the
normalized light curves, $S^N_{i,j}$, as as
$$
S^N_{i,j} = \frac{S_{i,j}}{<S_i>}, \qquad 
i = 1035, 1148, 1244, 1400, 1545, 1823, 1945
$$
where $S_{i,j}$ is the flux density of source $i$ on day $j$.  The
normalized light curves of the seven CSOs are shown in
Figure~\ref{fig_pltcal1}.  The $S^N$ curves show considerable scatter
about the mean value of 1.0, up to nearly 10\% for some days, and the
RMS scatter about the mean for the entire sample is 1.3\%.  This is a
little better than the canonical absolute accuracy of 2\% quoted for
standard VLA observations.  However, a significant portion of the
scatter is due to systematic rather than random errors.  Although it
may not be immediately obvious from the plot, the CSO light curves
track each other in the sense that on certain days all of the measured
flux densities will be above the mean and on other days all will be
below the mean.  This behavior results from small errors in the
absolute flux density calibration derived from 3C\,343, which are
due to a combination of errors in the AIPS calibration process (see
\S3) and the possible small intrinsic variability of the calibrator.
The same effect is seen, perhaps more clearly, in the steep-spectrum
source light curves shown in \citet{paper1}, for which the absolute
flux density calibration was derived from 3C\,286.  We have to remove
the calibration errors in order to evaluate properly the variability
of the CSO sample.

\begin{figure}
\plotone{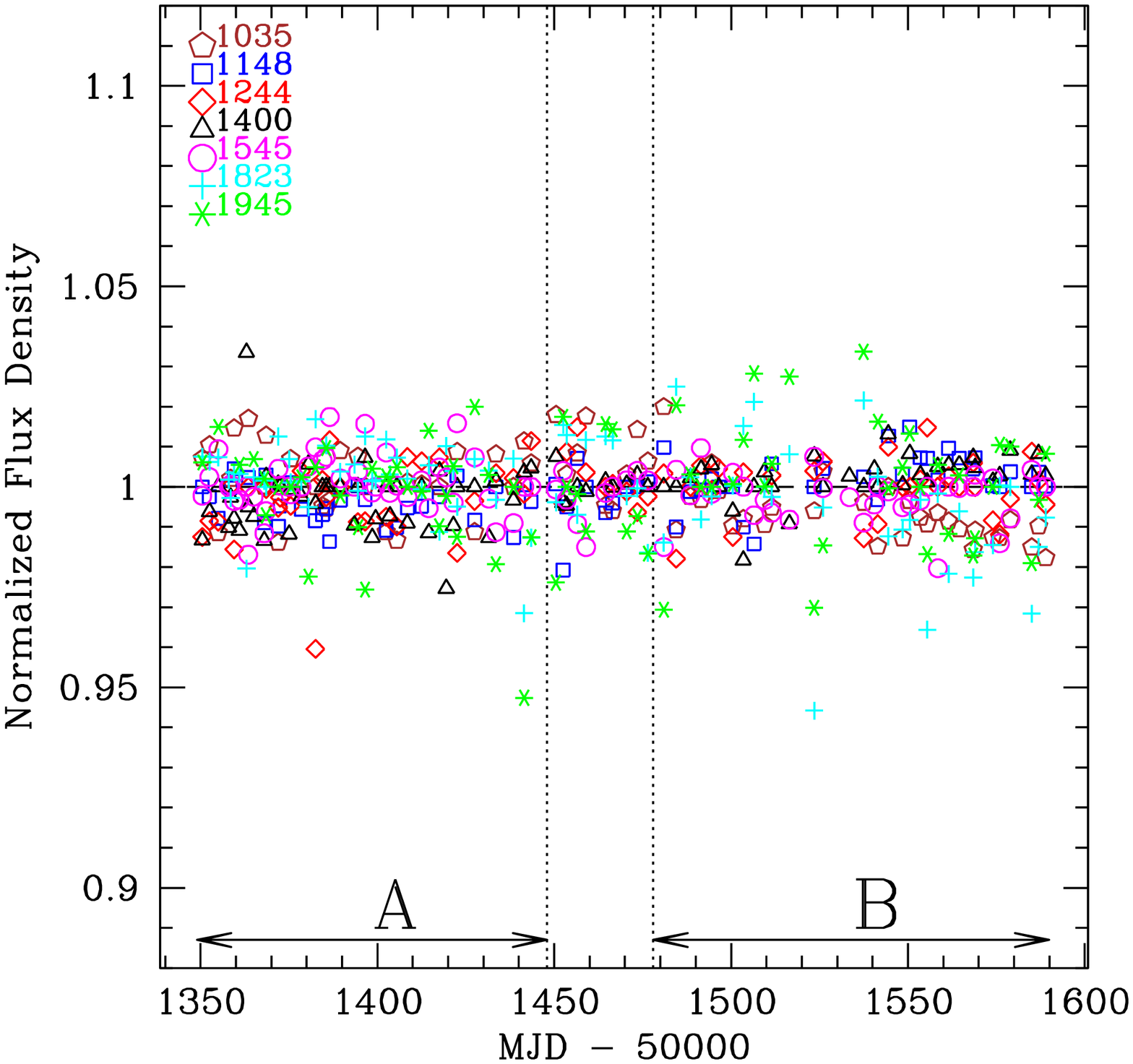}
\caption{Same as Figure~\ref{fig_pltcal1} except all light curves
have been divided by the median curve, $M^{(1)}$, to account for
errors in the flux density calibration.
\label{fig_pltcal2}}
\end{figure}

Fortunately, the data themselves contain the information necessary to
correct for the systematic errors in the flux density calibration.  We
assume that any intrinsic variability in our sample of CSOs is not
correlated from source to source.  Thus, any correlated variations in
the light curves, such as the large variations seen on days 1481 and
1523.5 in Figure~\ref{fig_pltcal1}, must be due to systematic errors
in the absolute flux density calibration.  To correct for the flux
calibration errors, we construct a light curve, $M^{(1)}$, that
represents our estimate of the effect of those errors.  This curve is
simply the median of the normalized CSO flux densities recorded for
each epoch, i.e.,
$$
M^{(1)}_j = {\rm median}(S^N_{i,j}).
$$
The resulting $M^{(1)}$ curve is represented by the dark line
in Figure~\ref{fig_pltcal1}.  We then divide the CSO light curves by
the median curve to perform the correction.  The median-corrected
light curves,
$$
S^M_{i,j} = \frac{S^N{i,j}}{M^{(1)}_j},
$$
are shown in Figure~\ref{fig_pltcal2}.  The variability in the
corrected light curves is now due to small, uncorrelated variations
about their mean values.  The RMS offsets from the mean are given in
the ``$M^{(1)}$ RMS'' column of Table~\ref{tab_rms}.  The total RMS of
the sample as a whole is 0.99\%.

\begin{deluxetable}{lrrccc}
\tablenum{3}
\tablewidth{0pt}
\scriptsize
\tablecaption{CSO Variability\label{tab_rms}}
\tablehead{
 &
 & \colhead{$<S_i>$}
 & \colhead{$M^{(1)}$ RMS}
 & \colhead{$M^{(2)}$ RMS}
 & \colhead{Intrinsic RMS}
\\
\colhead{Source}
 & \colhead{$N_i$}
 & \colhead{(mJy)}
 & \colhead{(\%)}
 & \colhead{(\%)}
 & \colhead{(\%)}
}
\startdata
J1035+5628 & 69 &  768 & 0.95    & 0.83    & 0.98    \\
J1148+5924 & 70 &  436 & 0.66    & 0.53    & 0.53    \\
J1244+4048 & 68 &  438 & 0.87    & 0.72    & 0.82    \\
J1400+6210 & 88 & 1108 & 0.75    & 0.50    & 0.48    \\
J1545+4751 & 87 &  323 & 0.67    & 0.75    & 0.86    \\
J1823+7938 & 81 &  537 & 1.3\phn & \nodata & \nodata \\
J1945+7055 & 73 &  441 & 1.5\phn & \nodata & \nodata \\
\enddata
\end{deluxetable}


\section{Secondary Systematic Errors \label{sec_second}}

\begin{figure}
\plotone{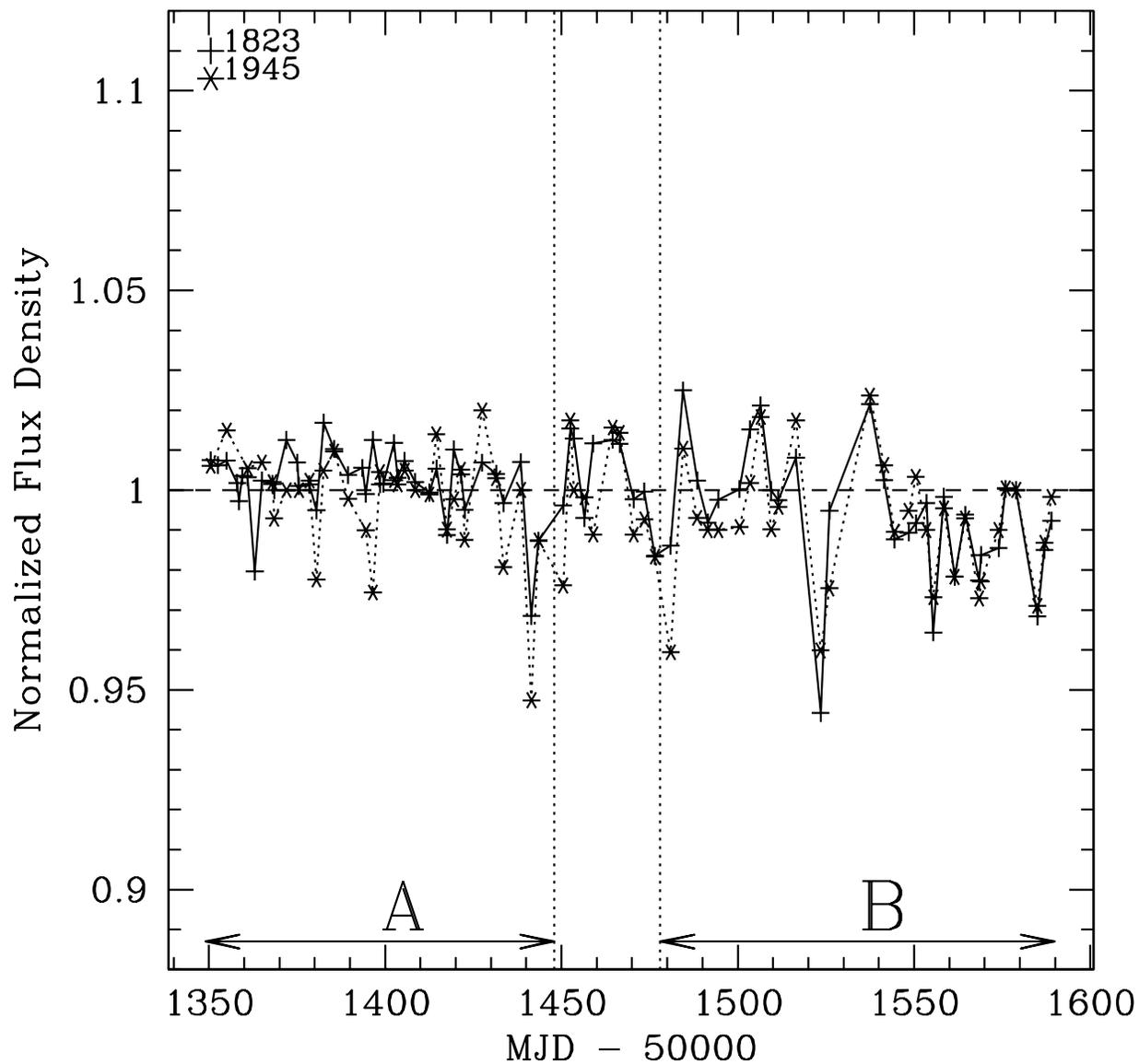}
\caption{Same as Figure~\ref{fig_pltcal2} except that only the light
curves for the sources J1823+7938 and J1945+7055 are shown.  The
variations in the two light curves appear correlated from day 1405
onward, and are especially well-matched from day 1540 onward.  To
emphasize the similarity of the day-to-day variations, the
B-configuration light curve for J1945+7055 has been shifted downward by
0.01.
\label{fig_1823}}
\end{figure}

Table~\ref{tab_rms} shows that the sources J1823+7938 and J1945+7055
have nearly double the RMS scatter of the other CSOs in the sample.
This can also be seen in the plot in Figure~\ref{fig_pltcal2} where
the J1823+7938 (+) and J1945+7055 (*) points are clearly visible above
the sample scatter, especially in the B-configuration data.  On the
face of it, these variations are small and would not be a cause for
worry.  However, upon closer examination of the light curves, it
becomes apparent that there is an additional systematic error
affecting the light curves of these two sources.  The presence of the
additional error is also clearly evident when we calculate the linear
correlation coefficient \citep[e.g.,][]{bevington} between all the
possible pairs of the CSO light curves.  We show the effect of this
error in Figure~\ref{fig_1823}, in which only the light curves of
J1823+7938 and J1945+7055, i.e., $S^M_{1823}$ and $S^M_{1945}$, are
plotted.  Not only do the light curves show a larger scatter than the
other CSO curves, but the deviations from the mean appear to be
correlated.  The light curves track each other very closely,
especially in the large variations seen in the B-configuration data.
We have not been able to determine the cause of this secondary
systematic error.  For example, the flux densities measured for these
two sources are not correlated with either the elevation of the CSO at
the time of observation, or the difference in elevation between the
flux calibrator (3C\,343) and the CSO.  We do note that these CSOs are
the two most northern in the sample, both with $\delta > 70^{\circ}$,
but we do not know of any property of the VLA that would only affect
far northern sources.  We have considered the possibility that this
error could be due to a systematic problem with the pointing model,
but we have found no other evidence in the long history of VLA
pointing runs to support this conclusion.  Because we cannot determine
the source of this systematic effect, we do not correct for it.  As a
result, the RMS variations listed in Table~\ref{tab_rms} for these two
sources should be regarded as upper limits to their true intrinsic
level of variability.  For the remainder of our discussion of the CSO
sample we will discard J1823+7938 and J1945+7055.

Although it is unlikely that systematic errors with the pointing
models can account for the correlated behavior of the light curves for
J1823+7938 and J1945+7055, it may be that small pointing errors are
introducing some of the scatter observed in the CSO light curves.  It
is possible to calculate corrections to the VLA antenna pointing
models through specialized observations of strong compact sources
prior to observing the sources of interest.  However, these
corrections need to be re-calculated each time the antennas make a
significant move.  Due to the wide spacing of our CSO sample on the
sky and the short observing blocks, we did not have sufficient time to
make the additional observations necessary to calculate these
corrections.  Future programs to study CSO variability should strongly
consider allocating time in their programs specifically for
determining the pointing solution corrections.  It should be possible
to use observations of the CSOs themselves to determine the
corrections.


\section{Final Light Curves and a Prescription for High-Precision
Radio Monitoring \label{sec_final}}

\begin{figure}
\plotone{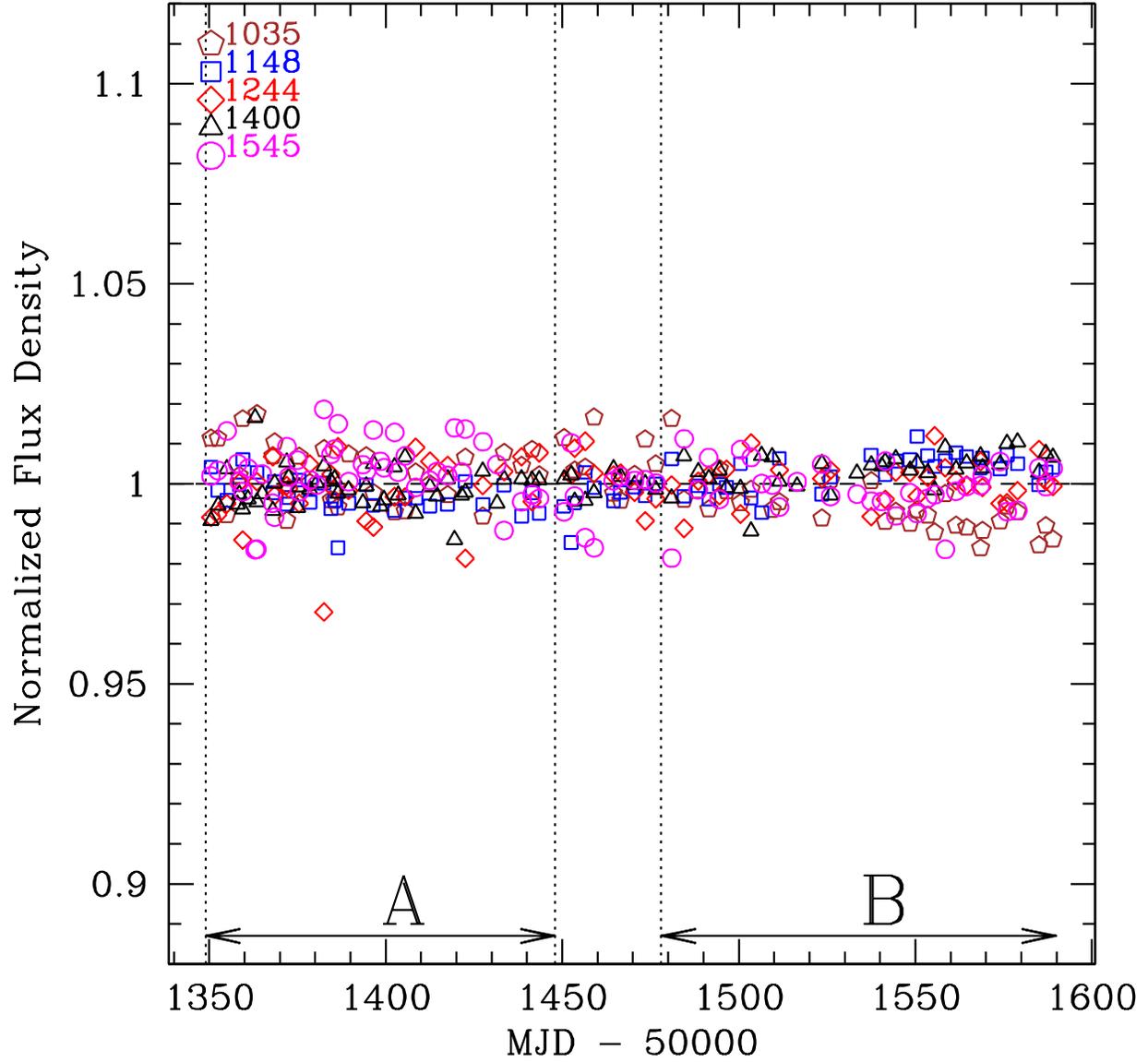}
\caption{Final CSO light curves.  The light curves have been divided
by the $M^{(2)}$ curve, which was constructed by excluding the
data from J1823+7938 and J1945+7055.
\label{fig_pltcal3}}
\end{figure}

To create the final CSO light curves, we discarded the data from
J1823+7938 and J1945+7055 and then used the light curves of the
remaining five CSOs to create a new correction curve, $M^{(2)}$.
Because the sources contributing the bulk of the outlier points had
been discarded, we created the $M^{(2)}$ curve from the mean of the
normalized light curves rather than their median, i.e.,
$$
M^{(2)}_j = \frac{\sum_i(S^N_{i,j})}{N_j}, \qquad
i = 1035,1148,1244,1400,1545
$$
where $N_j$ is the number of CSOs that were observed on day $j$.  The
final corrected light curves, 
$$
S^{\rm final}_{i,j} = \frac{S^N_{i,j}}{M^{(2)}_j}
$$
are shown in Figure~\ref{fig_pltcal3}.

It is now possible to estimate the intrinsic variability of the CSOs
from the final light curves.  In principle, the RMS scatter of the
$S^{\rm final}$ curves gives an estimate of the variability of the
sources.  However, it is important to realize that the division of the
input $S^N$ curves by the $M^{(2)}$ light curve is an operation on two
correlated quantities.  That is, each point in the $S^N$ curves gets
scaled by a quantity that includes a contribution from itself.  As a
result, the division by the $M^{(2)}$ curve has the effect, on
average, of depressing the measured RMS of the $S^{\rm final}$ curves
below their true intrinsic scatter.  We have estimated the correction
for this effect, in a statistical sense, by using the measured RMS
scatter in the $S^{\rm final}$ curves.  The measured and corrected RMS
scatters for each source are given in the ``$M^{(2)}$ RMS'' and
``Intrinsic RMS'' columns of Table~\ref{tab_rms}, respectively.  The
maximum correction is only $\Delta$\,RMS=$+$0.15\% and, thus, our
conclusion that the CSOs have extremely stable flux densities does not
change.  Whether we use the ``$M^{(2)}$'' or ``Intrinsic'' values for
the RMS variability of the CSOs, we find that the mean RMS scatter for
the five CSOs is $\sim$0.7\%.  This is the same value obtained by
considering the combined sample of CSOs to be one distribution and
calculating its RMS scatter about the mean.  Finally, we note that the
estimated intrinsic variabilities that we have derived above may still
include unknown systematic errors and thus should be regarded as upper
limits.

\begin{figure}
\plotone{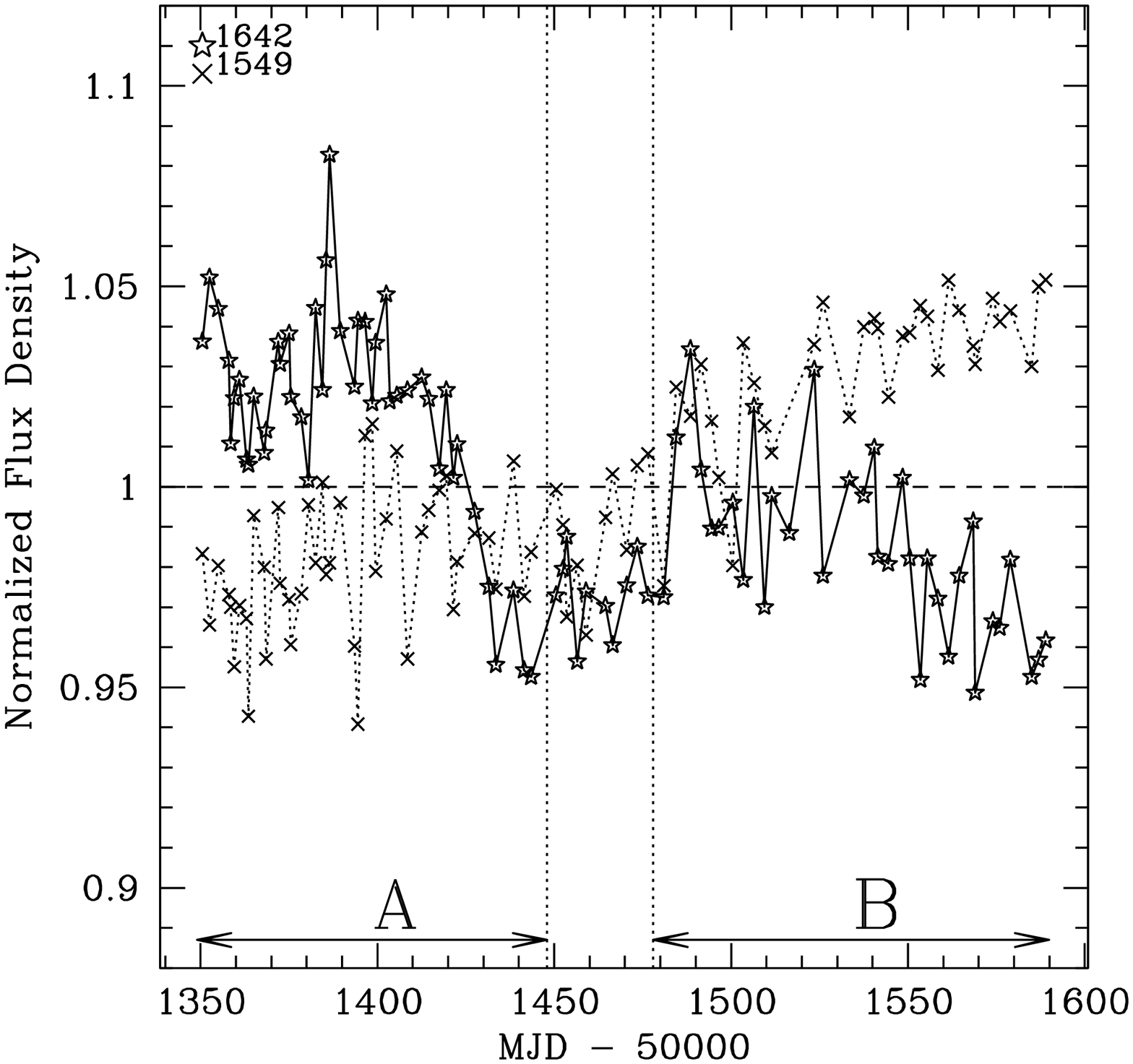}
\caption{Final light curves for the two core-jet sources.  These light
curves have been normalized by their mean flux density and then divided
by the $M^{(2)}$ curve.  Note that
both sources undergo systematic changes in flux density with time,
at a level of a few percent.
\label{fig_pltpcal}}
\end{figure}

The CSO light curves can be contrasted with those of the phase
calibrators, both of which are core-jet sources on the parsec scale
\citep{fey97}, shown in Figure~\ref{fig_pltpcal}.  The core-jet light
curves have been processed in the same manner as the CSO light curves,
including the correction with the $M^{(2)}$ curve.  Not only do the
core-jet calibrator curves have a higher RMS scatter than do the CSO
light curves (Table~\ref{tab_pcrms}), but the nature of the core-jet
light curves is qualitatively different from those of the CSOs.
Instead of random scatter about the mean value, the core-jet sources
show systematic changes in flux density during the course of the
observations, indicating true variability at the level of nearly 10\%.
They also exhibit variability from 2--5\% from epoch to epoch.

\begin{deluxetable}{cccc}
\tablenum{4}
\tablewidth{0pt}
\scriptsize
\tablecaption{Core-Jet Variability\label{tab_pcrms}}
\tablehead{
 &
 & \colhead{$<S_i>$}
 & \colhead{$M^{(2)}$ RMS}
\\
\colhead{Source}
 & \colhead{$N_i$}
 & \colhead{(mJy)}
 & \colhead{(\%)}
}
\startdata
J1549+5038 & 86 &  921 & 3.0 \\
J1642+6856 & 88 &  859 & 2.9 \\
\enddata
\end{deluxetable}

Given the nature of the CSO light curves, which show only small and
random variations about a mean value, we conclude that the CSO sources
should make excellent flux density calibrators for radio monitoring
campaigns.  If high-precision relative calibration is desired, we
recommend that at least four CSOs be observed in each epoch, if
possible.  With this observing strategy, it should be possible to
account correctly for systematic errors in the overall flux density
calibration.  In addition, having at least four CSO calibrators
enables the rejection of any one of the calibrators if something
adversely affects its measured flux density, for either the entire
monitoring campaign or a single epoch.  We emphasize, again, that CSOs
should not be equated with GPS sources, especially for the purpose of
flux density calibration.


\section{Physical Constraints on CSOs from Variability}

The average variability for five CSOs is 0.7\%, excluding the two
high-declination sources J1823+7938 and J1945+7055.  This average value
doubtless includes some, as yet poorly understood, residual systematic
errors.  Until such a time as we can demonstrate reproducible flux
density measurement with less scatter, this value should be considered
an upper limit on the true variability of CSOs.  Even so we can use
this limit to estimate the fraction of the luminosity coming from a
compact core, and eventually to constrain the evolution of CSOs.

The total variability of core-jet sources is typically $\sim$3\%
\citep[also see Figure~\ref{fig_pltpcal}]{qui92}.  Taking an average
core fraction of 0.6 from VLBI observations \citep{tay1996}, we can
estimate the real variability of the core components to be $\sim$5\%.
If the core components in CSOs behave similarly then we could use the
observed variability of 0.7\% to estimate that the core fraction in
CSOs to be $\sim$0.14.  In fact a more typical estimate from VLBI
images is 0.03 \citep{pinpoint}.  Both the CSO variability and the typical
core fraction are derived from small samples with only one source in
common (J1400+6210).  However, assuming that they are still indicative
of the average properties we can conclude that the core components in
CSOs are a factor 5 less variable than cores in core-jet sources.
This could indicate that the less-beamed cores are physically larger
in CSOs and therefore much less affected by refractive interstellar
scattering in the ISM.


\citet{acr96} derive empirical relationships for the evolution of of
powerful radio galaxies from the CSO phase into classical double radio
galaxies like Cygnus A.  They find evidence for a modest decrease in
luminosity, $L$, as the radio source grows in size, $r$, according to
$L \propto r^{q}$ where $q = -0.35$.  Current estimates of the hot
spot advance speed in CSOs is 0.1c \citep{ows98,tay2000}.  In 1 year a
typical CSO, of size $\sim$100~ly, will grow by 0.2~ly, or 0.2\%.  The
predicted change in luminosity is $-$0.07\%~yr$^{-1}$.  Our current
measurements range from $-$2.6 to 1.3\%~yr$^{-1}$, with an average
value of $-$0.42\%~yr$^{-1}$ and an RMS of 1.7\%~yr$^{-1}$.  To
provide an adequate test of the empirical relations proposed by
\citet{acr96} will require an order of magnitude improvement in the
measured variations.  Observations of the sort reported here, spread
out over $\sim$200 days are unlikely to achieve this with currently
available instruments.  With fiber optic transmission, a new
correlator, and additional antennas in the southwest, the Expanded VLA
(EVLA) should have both increased stability and the capability of
using a subarray comprised of antennas in the southwest US to monitor
gravitational lenses and CSOs with high resolution for the $\sim$3
years needed to achieve the desired accuracy.


\section{Summary}

We monitored a sample of seven CSOs with the VLA over the course of
eight months.  The flux densities of all of the sources were very
stable, with only small measured variations about their mean value.
We find that five of the CSOs have RMS variations of less than 1\%.
For these sources the average measured RMS is 0.7\% and the RMS
variability of the sample as a whole is also 0.7\%.  For the remaining
two CSOs we estimate upper limits on the RMS variability of 1.5\% or
less.  We conclude that CSOs, in general, are flux stable on time
scales between 1 week and 10 months.  Given this stability, they are
excellent flux density calibrators for monitoring experiments.  For a
high precision experiment, we recommend that at least four of these
CSOs be included in the observations.


\acknowledgments For useful discussions, we thank Neal Miller, Stefano
Casertano, Tim Pearson, Alison Peck, Lori Lubin, Steve Myers, Frazer
Owen, Rick Perley, Adam Riess, and Ken Sowinski.  We are grateful to
Leon Koopmans, David Rusin, and Emily Xanthopoulos for help in
choosing the CSO sample and in planning the observations.  We thank
the anonymous referee for suggestions that improved the paper.  Meri
Stanley, Jason Wurnig, and Ken Hartley did their usual excellent job
in checking over the observing schedules and catching mistakes.  We
thank the NRAO staff for keeping the VLA running smoothly.



\clearpage

\begin{thebibliography}{}

\bibitem[Aller, Hughes, \& Aller(1990)]{aller90} Aller, H.\ D.\, Hughes, 
P.\ A., \& Aller, M.\ F.\ 1990, in ``Variability of Active Galactic 
Nuclei'' eds. R.\ Miller \& P.\ J.\ Wiita, (Cambridge Univ. Press:Cambridge),
p. 172

\bibitem[Baum et al.(1990)]{bau90} Baum, 
S.\ A., O'Dea, C.\ P., de Bruyn, A.\ G., \& Murphy, D.\ W.\ 1990, \aap, 
232, 19 

\bibitem[Bevington(1969)]{bevington}Bevington, P.\ R. 1969, {\em
Data Reduction and Error Analysis for the Physical Sciences},
(New York: McGraw-Hill)

\bibitem[de Bruyn(1991)]{debruyn} de Bruyn, A.\ G.\ 1991, 
in {\it Variability of Active Galaxies}, eds. W.J.\ Duschl, S.J.\ Wagner
\& M. Camenzind, (Springer--Verlag: New York) p.\ 105 

\bibitem[Fassnacht et al.(1999)]{paper1} Fassnacht, C. D., Pearson, T. J., 
Readhead, A. C. S., Browne, I. W. A., Koopmans, L. V. E., Myers, S. T., 
\& Wilkinson, P. N. 1999, \apj, 527, 498

\bibitem[Fey \& Charlot(1997)]{fey97} Fey, A.\ L.\ \& 
Charlot, P.\ 1997, \apjs, 111, 95 

\bibitem[Owsianik \& Conway(1998)]{ows98} Owsianik, I.\ \& 
Conway, J.\ E.\ 1998, \aap, 337, 69 

\bibitem[Owsianik, Conway, \& Polatidis(1998)]{owspol98} 
Owsianik, I., Conway, J.\ E., \& Polatidis, A.\ G.\ 1998, \aap, 336, L37 

\bibitem[Peck \& Taylor(2000)]{csosamp} Peck, A. B. \& Taylor, G. B. 2000,
\apj, 534, 90

\bibitem[Phillips \& Mutel(1980)]{phi80} Phillips, R.\ B.\ \& 
Mutel, R.\ L.\ 1980, \apj, 236, 89 

\bibitem[Phillips \& Mutel(1982)]{phi82} Phillips, R.\ B.\ \& 
Mutel, R.\ L.\ 1982, \aap, 106, 21 

\bibitem[Quirrenbach et al.(1992)]{qui92} Quirrenbach, A.\ et 
al.\ 1992, \aap, 258, 279 

\bibitem[Readhead et al.(1996)]{rea96} Readhead, A.\ C.\ S., 
Taylor, G.\ B., Xu, W., Pearson, T.\ J., Wilkinson, P.\ N., \& Polatidis, 
A.\ G.\ 1996, \apj, 460, 612 

\bibitem[Readhead et al.(1996)]{acr96} Readhead, A.\ C.\ S., Taylor, G.\ B., 
Pearson, T.\ J., \& Wilkinson, P.\ N.\ 1996, \apj, 460, 634 

\bibitem[Rudnick \& Jones(1982)]{rud82} Rudnick, L.\ \& 
Jones, T.\ W.\ 1982, \apj, 255, 39 

\bibitem[Shepherd(1997)]{difmap} Shepherd, M.\ C. 1997, in 
{\em Astronomical Data Analysis Software and Systems VI}, eds. 
G.\ Hunt \& H.\ E.\ Payne, (ASP Conference Series, v125) 77

\bibitem[Snellen, Schilizzi, \& van Langevelde(2000)]{snel00} Snellen,
I. A. G., Schilizzi, R. T., \& van Langevelde, H. J. 2000, \mnras,
319, 429

\bibitem[Stanghellini et al.(1997)]{sta97} Stanghellini, C., 
Bondi, M., Dallacasa, D., O'Dea, C.\ P., Baum, S.\ A., Fanti, R., \& Fanti, 
C.\ 1997, \aap, 318, 376 

\bibitem[Taylor et al.(1996a)Taylor, Readhead, \& Pearson]{pinpoint} 
Taylor, G.\ B., Readhead, A.\ C.\ S., \& Pearson, T.\ J.\ 1996a, \apj, 463, 95 

\bibitem[Taylor et al.(1996b)]{tay1996} Taylor, G.\ B., 
Vermeulen, R.\ C., Readhead, A.\ C.\ S., Pearson, T.\ J., Henstock, D.\ R., 
\& Wilkinson, P.\ N.\ 1996b, \apjs, 107, 37 

\bibitem[Taylor, Wrobel, \& Vermeulen(1998)]{tay98} 
Taylor, G.\ B., Wrobel, J.\ M., \& Vermeulen, R.\ C.\
1998, \apj, 498, 619 

\bibitem[Taylor et al.(2000)]{tay2000} 
Taylor, G.\ B., Marr, J.\ M., Pearson, T.\ J., \& Readhead, A.\ C.\ S.\ 
2000, \apj, 541, 112 

\bibitem[van Gorkom et al.(1989)]{jvg89} van Gorkom, J.\ H.,
Knapp, G.\ R., Ekers, R.\ D., Ekers, D.\ D., Laing, R.\ A., 
\& Polk, K.\ S.\ 1989, \aj, 97, 708

\bibitem[Vermeulen \& Cohen(1994)]{coh94} Vermeulen, R.\ C.\ 
\& Cohen, M.\ H.\ 1994, \apj, 430, 467 

\bibitem[Wilkinson et al.(1994)]{wil94} Wilkinson, P.\ N., 
Polatidis, A.\ G., Readhead, A.\ C.\ S., Xu, W., \& Pearson, T.\ J.\ 1994, 
\apjl, 432, L87 

\bibitem[Zhou et al.(2000)]{zhou2000} Zhou, 
J.\ F., Hong, X.\ Y., Jiang, D.\ R., \& Venturi, T.\ 2000, \apjl, 541, L13 

\end{thebibliography}
\end{document}